\documentstyle[psfig,preprint,aps]{revtex}

\tightenlines

\def\beq{\begin{equation}}
\def\eeq{\end{equation}}
\def\bea{\begin{eqnarray}}
\def\eea{\end{eqnarray}}

\begin{document}
\hoffset-1cm
\draft

\title{Soft Thermal Loops in Scalar Quantum Electrodynamics
\footnote{Supported by BMBF, GSI Darmstadt, and DFG}}

\author{Stefan Leupold and Markus H. Thoma\footnote{Heisenberg fellow}}
\address{Institut f\"ur Theoretische Physik, Universit\"at Giessen, 35392 
Giessen, Germany}

\date{\today}

\maketitle

\begin{abstract}
Scalar QED is used to study approximations in thermal field theory 
beyond the Hard Thermal Loop resummation scheme. For this purpose
the photon self energy is calculated for external momenta of the
order $e^2T$ using Hard Thermal Loop resummed Green functions.
\end{abstract} 

\bigskip

{\hspace*{1.1cm}Keywords: thermal field theory, Hard Thermal Loop
resummation}

\medskip


\narrowtext
\newpage

During the last years thermal field theory became increasingly important 
for investigating problems in nuclear physics and astrophysics. 
Famous examples are the prediction of signatures for the quark-gluon plasma 
formation in relativistic heavy ion collisions \cite{ref1}, the
origin of the baryon asymmtery in the early universe \cite{ref2},
and neutrino induced reactions in supernovae \cite{ref3}. 

Perturbative calculations based on bare propagators and vertices, however,
turned out to be inconsistent, i.e., they lead to gauge dependent and
infrared divergent results. The reason for this failure is that 
at finite temperature (and density) higher order loop diagrams can contribute 
to lower order in the coupling constant. This problem has been solved
within the Hard Thermal Loop (HTL) resummation technique invented
by Braaten and Pisarski \cite{ref4}. The basic idea is that one has
to distinguish between momentum scales $T$, called hard, and $gT$, called
soft, in the weak coupling limit $g\ll 1$. Now, the leading order contribution
to self energies and vertices with soft external momenta comes from 
one loop diagrams containing only hard momenta. By resumming these HTLs, 
effective propagators and vertices are constructed containing medium effects
such as Debye screening. These effective Green 
functions have to be used if their momenta are soft, while bare Green 
functions are sufficient for hard momenta. In this way, quantities,
in which only scales of the order $T$ and $gT$ are involved, can be treated
consistently. 

However, problems being sensitive to the scale $g^2T$, such as 
damping rates of hard quarks and gluons in the quark-gluon plasma
\cite{ref5} and the sphaleron rate possibly related to baryogenesis
in the early universe \cite{ref6}, cannot be treated in this way. 
The super soft scale\footnote{Sometimes the scale $g^2T$ is called
soft and the scale $gT$ semi hard} $g^2T$ arises from the transverse 
part of the massless gauge boson propagator, which shows in the HTL 
approximation no screening in the static 
limit, i.e. no screening of static magnetic fields. Therefore quantities 
involving the transverse gauge boson propagator at scales $g^2T$ or smaller 
exhibit infrared singularities, although HTL resummed propagators are used.
For example, the damping rate of hard partons calculated
within the HTL resummation method suffers from a logarithmic
infrared divergence \cite{ref5}.

For dealing with the momentum scale $g^2T$, one has to go beyond
the HTL approximation. For static quantities, such as the thermodynamic
quantities entropy, free energy, energy density, and pressure,
an approximation beyond the HTL method based on an effective high 
temperature field theory has been derived using dimensional reduction
\cite{ref7}. For dynamical quantities, e.g. particle production and 
interaction rates, the super soft scale has been considered in the 
non-abelian case using a Langevin equation for the gauge fields
\cite{ref8}. This Langevin equation is related to the self energies
of the gauge boson at super soft external momentum \cite{ref9}.
In order to calculate this self energy one has to consider
diagrams containing soft momenta, which we will call Soft Thermal Loops
(STL) in the following. The explicit evaluation of these diagrams is very 
involved, because one has to use HTL propagators and vertices. As an
example we show the STL photon self energy in Fig.1, where blobs denote
HTL Green functions. 

In the case of scalar QED, however, STLs can be computed easily and
analytic expressions can even be given as we will see in the following. 
Calculations are facilitated for scalar QED significantly by the
fact that HTL effective vertices are absent and the HTL scalar
self energy is momentum independent \cite{ref10}. After all scalar QED is 
a gauge theory which shows a sensitivity to the scale $g^2T$ in the gauge 
boson sector, i.e., it suffers from the absence of static magnetic
screening in the same way as QED and non-abelian gauge theories.
Therefore scalar QED has been used as a toy model in a number of 
papers for studying problems in finite temperature gauge field theories
\cite{ref10,ref11}. A possible application of scalar QED at finite temperature
might be the formation of topological defects in the early universe 
\cite{ref12}.
Also it can be easily extended to the Higgs sector of the electroweak
theory in the symmetric phase \cite{ref13}.

The aim of the present investigation is the computation of the STL
photon self energy in scalar QED, which is the basic ingredient for
going beyond the HTL resummation scheme. 

Since there are no HTL vertices in scalar QED, the STL photon self energy is 
given by the diagrams shown in Fig.2. Using the real time formalism
within the Keldysh representation (see e.g. \cite{ref13a}), the retarded,
advanced, and symmetric scalar HTL propagator, assuming a vanishing 
bare scalar mass, is given by 
\bea
\Delta_{R,A}^*(K)&=&\frac{1}{K^2-m_s^2\pm i sgn(k_0) \epsilon}\, ,\nonumber \\
\Delta_F^*(K)&=&-2\pi i\> [1+2n_B(|k_0|)]\> \delta (K^2-m_s^2)\, ,  
\label{eq1}
\eea
where we have used the notation $K=(k_0, {\bf k})$, $k=|{\bf k}|$, and
$n_B(x)=1/[\exp(x/T)-1]$ denotes the Bose distribution function.
The effective, temperature dependent scalar mass is given by $m_s=eT/2$
\cite{ref10}. Although the one-loop scalar self energy contains a polarization
diagram with two internal photon lines, it reduces to a momentum independent
mass term in the HTL approximation. Similar there are no effective HTL 
vertices in the HTL approximation, which can be traced back to the absence
of $\gamma $-matrices in the bare vertices. Therefore the STL photon self 
energy differs from the HTL self energy only by the presence of the
effective scalar mass $m_s$. 

Let us first discuss the tadpole contribution of Fig.2. Using the real time
formalism the retarded self energy contribution can be written as
\beq
^{tad}{\Pi_R^*}^{\mu \nu}=-ie^2\> g^{\mu \nu}\> \int \frac{d^4K}{(2\pi )^4}\>
[\Delta_F^*(K)+\Delta_R^*(K)+\Delta_A^*(K)]\, .
\label{eq2}
\eeq
Now, assuming the external legs of this self energy to be super soft,
we want to extract the leading order correction to the HTL self energy. This
correction is of order $e^3$ and comes entirely from the part containing
the distribution function in (\ref{eq2}) \cite{ref14}. The terms without 
distribution functions lead to a correction of order $e^4$ and will be 
neglected \cite{ref5}. After integrating over $k_0$ by means of
the $\delta$-functions in (\ref{eq1}) and trivially over the angles,
(\ref{eq2}) reduces to 
\beq
^{tad}{\Pi_R^*}^{\mu \nu}\simeq -\frac{e^2}{\pi ^2}\> g^{\mu \nu}\> 
\int_0^\infty dk\> \frac{k^2}{\omega _k}\> n_B(\omega _k)\, ,
\label{eq3}
\eeq
where $\omega_k^2=k^2+m_s^2$. The order $e^3$ contribution can be extracted
by introducing a separation scale $eT\ll \Lambda \ll T$. Decomposing
the integral in (\ref{eq3}) in two parts, where we integrate in the first
one from zero to $\Lambda$ and in the second one from $\Lambda$ to infinity,
we may expand the distribution function in the first integral for small
arguments and neglect $m_s$ of order $eT$ in the second. Then we end
up with
\beq
^{tad}{\Pi_R^*}^{\mu \nu}= -\frac{3}{2}\>  g^{\mu \nu}\> m_\gamma ^2\> 
\left (1-\frac{3}{2\pi} e\right ) + O(e^4)\, ,
\label{eq4}
\eeq
where $m_\gamma =eT/3$ is identical to the plasma frequency in the HTL limit.
The HTL result is given by the first term proportional to $e^2$
in (\ref{eq4}).  

For calculating the polarization diagram of Fig.2 it is convenient to
decompose it into a longitudinal and transverse part\footnote{At finite 
temperature the photon self energy has two independent components, for 
which we can choose the longitudinal part $\Pi_L=\Pi ^{00}$ and the 
transverse part $\Pi_T=(\delta _{ij}-p_ip_j/p^2)\Pi ^{ij}/2$.}. 
Using the real time formalism the second diagram of Fig.2 leads to
\beq
^{pol}{\Pi_R^*}^L=i\frac{e^2}{2}\> \int \frac{d^4K}{(2\pi )^4}\>
(q_0+k_0)^2\> 
[\Delta_F^*(Q)\Delta_R^*(K)+\Delta_A^*(Q)\Delta_F^*(K)]\, ,
\label{eq5}
\eeq
where $Q=K-P$. Since the internal scalar propagators are never super
soft due to the presence of the effective scalar mass of the order $eT$,
the integrands in (\ref{eq5}) can be expanded for super soft $p,p_0\ll
\omega _k$. This expansion is analogous to the one for
extracting the HTL contribution, where the internal momenta are assumed
to be hard, while the external are soft. Then (\ref{eq5}) yields
\beq
^{pol}{\Pi_R^*}^L\simeq \frac{e^2}{\pi ^2}\> \> \int_0^\infty 
dk\> k \> n_B(\omega _k)
\> \left (\frac{p_0}{p}\, \ln \frac{p_0\omega_k+pk+i\epsilon}{p_0\omega_k
-pk+i\epsilon}-P^2\, \frac{\omega_kk}{p_0^2\omega_k^2-p^2k^2}\right ).
\label{eq6}
\eeq
For $\omega_k=k$ this expression reduces to the HTL result
\beq
^{pol}\Pi_R^L=-\frac{e^2T^2}{6}\left (1-\frac{p_0}{p}\, \ln 
\frac{p_0+p+i\epsilon}{p_0-p+i\epsilon}\right ).
\label{eq7}
\eeq

Proceeding similar as for the tadpole by introducing a separation scale
$\Lambda$, the leading order correction follows from
\bea
^{pol}{\Pi_R^*}^L&&\simeq \frac{e^2T^2}{\pi^2}\> \Biggl \{\frac{\pi^2}{6}
\left (\frac{p_0}{p}\, \ln \frac{p_0+p}{p_0-p}-1\right )
\nonumber \\
&&-\frac{1}{T}\int_0^\infty dk\left [\frac{p_0}{p}\, \ln 
\frac{p_0+p}{p_0-p}-1-z\left(\frac{p_0}{p}\, \ln \frac{p_0+pz}{p_0-pz}
-P^2\, \frac{z}{p_0^2-p^2z^2}\right )\right ]\Biggr \}\, ,
\label{eq8}
\eea
where $z=k/\omega_k$ and we neglected $i\epsilon $. Using the substitution 
$k\rightarrow z$ the integral in (\ref{eq8}) can be evaluated analytically 
resulting in
\beq
{\Pi_R^*}^L=\Pi_R^L+\frac{e^3T^2}{4\pi }\left (1-\frac{p_0}{\sqrt{p_0^2-p^2}}
\right )+O(e^4)\, ,
\label{eq9}
\eeq
where the tadpole contribution (\ref{eq4}) has been added and
\beq
\Pi_R^L=-3\, m_\gamma^2\> \left (1-\frac{p_0}{2p}\, \ln 
\frac{p_0+p}{p_0-p}\right )
\label{eq10}
\eeq
is the longitudinal HTL self energy \cite{ref15}. 

Analogously one finds for the transverse part of the photon self energy
\beq
{\Pi_R^*}^T=\Pi_R^T+\frac{e^3T^2}{4\pi }\left [\left (1-\frac{p_0^2}{p^2}
\right )\> \left (1-\frac{p_0}{\sqrt{p_0^2-p^2}}\right )-1\right ] +O(e^4)\, ,
\label{eq11}
\eeq
where the transverse HTL self energy is given by \cite{ref15}
\beq
\Pi_R^T=\frac{3}{2}\, m_\gamma^2\, \frac{p_0^2}{p^2}\> \left [1-\left (1-
\frac{p^2}{p_0^2}\right )\,\frac{p_0}{2p}\, \ln 
\frac{p_0+p}{p_0-p}\right ]\, .
\label{eq12}
\eeq
Actually the STL transverse photon self energy is more interesting
because transverse photons are sensitive to the scale $e^2T$ due to
the missing static magnetic screening. The photon self energy 
(\ref{eq9}) and (\ref{eq11}) agrees with the expressions given in 
Ref.\cite{ref10} in the super soft momentum limit $p_0,p\sim e^2T$. 

Now we want to discuss our results. First we observe that
in contrast to non-abelian theories there is no STL correction to
order $e^2$ for super soft external momenta as already noted by B\"odeker 
\cite{ref9}. Secondly, we note that our result (\ref{eq9}) and (\ref{eq10})
are gauge invariant, since $m_s$ is a gauge invariant quantity.
Next, we see that the self energies given here possess an imaginary part
only below light cone $p_0^2<p^2$ as the HTL self energies do. However,
the imaginary part comes now from a square root instead of a logarithm.

Finally, we want to study the zero momentum limits of our 
self energies. In the limit $p\rightarrow 0$ the expressions
in (\ref{eq9}) and (\ref{eq11}) reduce to
\bea
{\Pi_R^*}^L(p=0)&=&\frac{p^2}{p_0^2}\, \left (m_\gamma ^2-\frac{e^3T^2}{8\pi}
\right )\, ,\nonumber \\
{\Pi_R^*}^T(p=0)&=&m_\gamma^2-\frac{e^3T^2}{8\pi }\, .
\label{eq13}
\eea
Hence the dielectric functions defined by $\epsilon _L=1-\Pi_L/p^2$
and $\epsilon _T=1-\Pi_T/p_0^2$ \cite{ref16} agree in this limit as expected,
since there is no direction preferred for $p=0$ \cite{ref17}.
The result (\ref{eq13}) should not be confused with the plasma frequency,
which is given by ${\Pi_R^*}^{L,T}(p_0=m_\gamma,p=0)$ \cite{ref10}.
This limit has not been studied here as we assumed that $p_0\sim e^2T$.
Also we cannot consider the plasmon dispersion relation from
our self energies (\ref{eq9}) and (\ref{eq11}), since $p_0\geq m_\gamma$
for real plasma modes.
In the limit $p_0\rightarrow 0$, on the other hand, the self energies read
\bea
{\Pi_R^*}^L(p_0=0)&=&-
3m_\gamma ^2\left (1-\frac{3e}{4\pi}
\right )\, ,\nonumber \\
{\Pi_R^*}^T(p_0=0)&=&\Pi_R^T(p_0=0)=0\, .
\label{eq14}
\eea
The result for the longitudinal component determines the correction to the
Debye screening mass, which is given by $m_D^2=3m_\gamma $ in the HTL
approximation\footnote{As in the HTL limit ${\Pi_R^*}^L (p_0=0,p)$ 
does not depend on $p$, thus allowing its interpretation as Debye mass 
\cite{ref18}.} and agrees with the one found 
in Ref.\cite{ref10,ref19}. 
As expected \cite{ref19} there is no static magnetic 
screening mass in scalar QED even beyond the HTL approximation.

\begin{figure}

\centerline{\psfig{figure=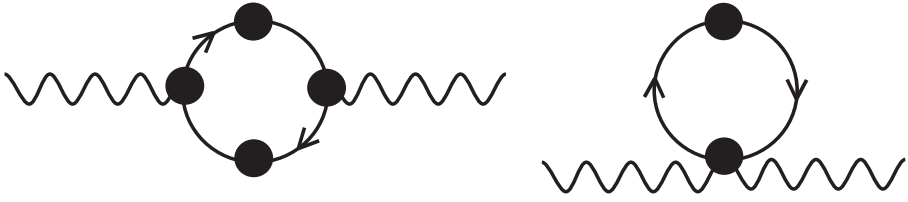,width=12cm}}
\vspace*{1cm}
\caption{STL photon self energy in QED. The blobs denote HTL resummed 
electron propagators and photon-electron vertices.}

\end{figure}

\vspace*{2cm}

\begin{figure}

\centerline{\psfig{figure=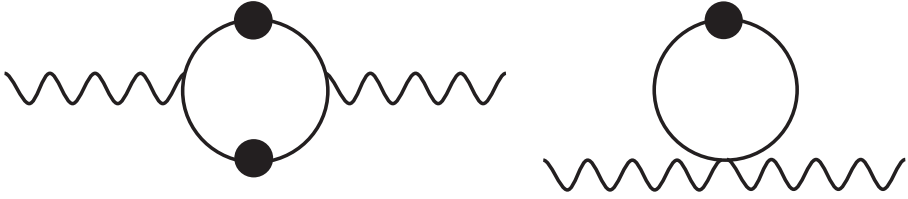,width=12cm}}
\vspace*{1cm}
\caption{STL photon self energy in scalar QED. The blobs denote HTL resummed
scalar propagators.}

\end{figure}

\end{document}